\documentclass[12pt]{article}
\usepackage{refmerge}
\usepackage{epsfig}
\begin{document}
\begin{center}
{\bf{ \boldmath
STUDY OF THE $\phi$ DECAYS INTO  
$\pi^{0}\pi^{0}\gamma$ AND $\eta\pi^{0}\gamma$ FINAL STATES
}}
\end{center}
\begin{center}
R.R.Akhmetshin, E.V.Anashkin, M.Arpagaus, V.M.Aulchenko,
V.S.Banzarov, L.M.Barkov, N.S.Bashtovoy, A.E.Bondar, D.V.Bondarev,
A.V.Bragin,
D.V.Chernyak, A.S.Dvoretsky,
S.I.Eidelman, G.V.Fedotovich, N.I.Gabyshev,
A.A.Grebeniuk, D.N.Grigoriev, P.M.Ivanov, S.V.Karpov, V.F.Kazanin,
B.I.Khazin, I.A.Koop, P.P.Krokovny,
L.M.Kurdadze, A.S.Kuzmin,
I.B.Logashenko, P.A.Lukin, A.P.Lysenko, K.Yu.Mikhailov, 
I.N.Nesterenko, V.S.Okhapkin, E.A.Perevedentsev, E.A.Panich, A.S.Popov, 
T.A.Purlatz, N.I.Root, A.A.Ruban, N.M.Ryskulov,
A.G.Shamov, Yu.M.Shatunov, B.A.Shwartz, A.L.Sibidanov, V.A.Sidorov,
A.N.Skrinsky, V.P.Smakhtin, I.G.Snopkov, E.P.Solodov, 
P.Yu.Stepanov, A.I.Sukhanov, 
V.M.Titov, Yu.V.Yudin, S.G.Zverev
\\
  Budker Institute of Nuclear Physics, Novosibirsk, 630090, Russia
\end{center}
\begin{center}
                J.A.Thompson
\\
                University of Pittsburgh, Pittsburgh, PA 15260, USA
\end{center}
\vspace{0.7cm}
\begin{abstract}
\hspace*{\parindent}
Radiative decays of the $\phi$ meson have been studied
using a data sample of about 19 million $\phi$ decays
collected by the CMD-2 detector at VEPP-2M collider in Novosibirsk.
From selected 
$e^+e^-\to\pi^{0}\pi^{0}\gamma$ and $e^+e^-\to\eta\pi^{0}\gamma$ 
events the following model independent results have been obtained:

 $Br(\phi\to\pi^{0}\pi^{0}\gamma) = (0.92\pm 0.08\pm0.06)\times10^{-4}$
for $M_{\pi^{0}\pi^{0}}>700$ MeV,

 $Br(\phi\to\eta\pi^{0}\gamma) = (0.90\pm 0.24\pm 0.10)\times10^{-4}$.\\
 It is shown that the intermediate mechanism $f_{0}(980)\gamma$ dominates 
in the $\phi\to\pi^{0}\pi^{0}\gamma$ decay and
the corresponding branching ratio is

 $Br(\phi\rightarrow f_{0}(980)\gamma)=(2.90\pm 0.21\pm1.54)\times10^{-4}$.\\
The systematic error is dominated by the possible model uncertainty.

Using the same data sample the upper limit has been obtained  
for the P- and CP-violating decay of $\eta$ at 90\% CL:

 $Br(\eta\rightarrow\pi^{0}\pi^{0}) < 4.3\times10^{-4}$ . 
\end{abstract}
\baselineskip=17pt
\section*{ \boldmath Introduction}
\hspace*{\parindent}
 In the preceding paper \cite{ppg} we reported on the first
observation of the $\phi\to\pi^+\pi^-\gamma$ decay with the CMD-2 detector.    
The present work provides complementary information on the 
electric dipole radiative transitions of the $\phi$ meson obtained from the
$\pi^{0}\pi^{0}\gamma$ and $\eta\pi^{0}\gamma$ final states.

The purely neutral decay  $\phi\to\pi^{0}\pi^{0}\gamma$ has no 
bremsstrahlung background as the $\phi\to\pi^+\pi^-\gamma$ decay
and is the most efficient to study the two pion mass spectrum in the 
$\phi\to f_{0}(980)\gamma$  decay.
With the CMD-2 detector this mode as well as another $\phi$ decay with five
photons in the final state  $\phi\to\eta\pi^{0}\gamma$ 
have been studied and the first results~\cite{ICHEP98} based on about 
25\% of the data confirmed those reported earlier by
SND~\cite{SNDa0g,SNDppg}.  

The  CMD-2  detector  described  in  detail  elsewhere~\cite{CMD285} has 
been taking data since 1992. 
In addition to the barrel CsI calorimeter, the endcap calorimeter 
made of BGO crystals was installed in 1995 making the detector almost 
hermetic to the photons.
The energy resolution for photons in the CsI calorimeter is about
 8\% independent of the energy and 
 $\sigma_{E}/E =4.6\%/\sqrt{E(GeV)}$ for the BGO calorimeter.

In total, the 14.2 $pb^{-1}$ of data have been collected  at
14 energy points around the $\phi$ mass. 
For the analysis of the $\phi\to\pi^{0}\pi^{0}\gamma$
and $\phi\to\eta\pi^{0}\gamma$ decays presented here 12.8 $pb^{-1}$ were 
used corresponding to  $18.8 \times 10^{6}$ $\phi$ decays.

%
\section*{\boldmath $\phi\to\pi^{0}\pi^{0}\gamma$ channel}
\subsection*{Selection of $\pi^{0}\pi^{0}\gamma$ Events}
\begin{figure}[tbh]
\vspace{-0.5cm}
\psfig{figure= 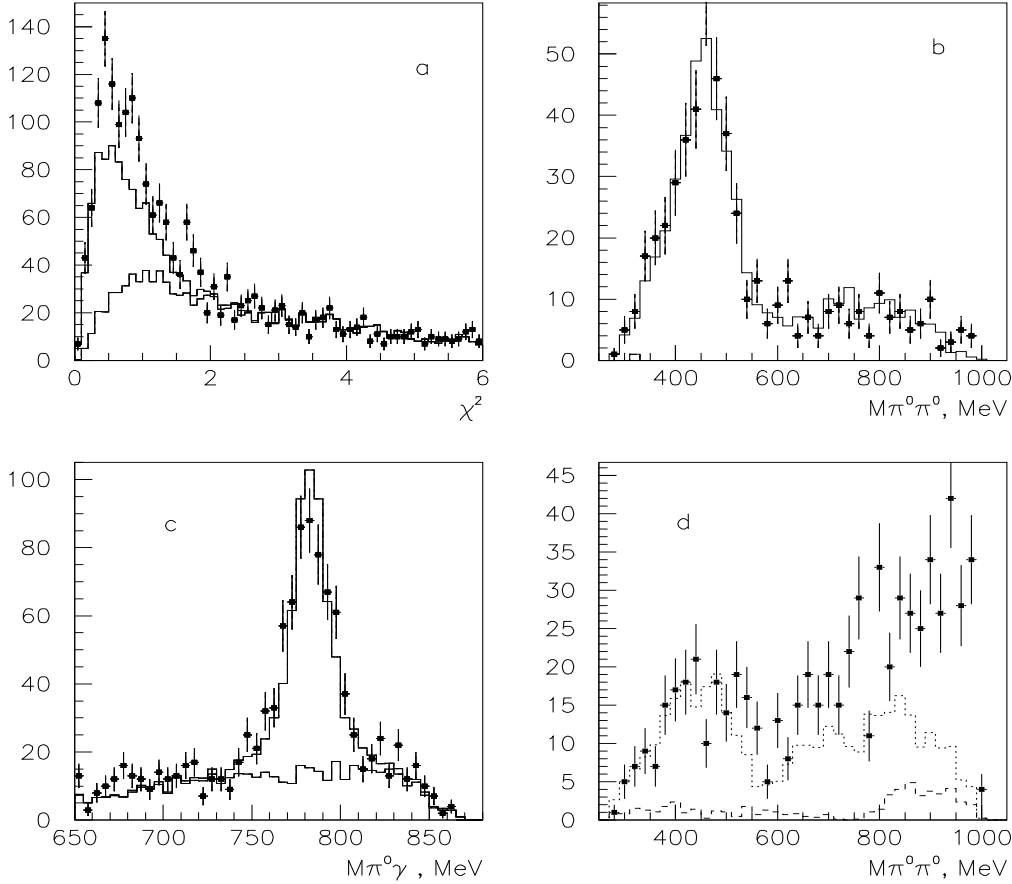, width=1.0\textwidth, height=0.8\textwidth}
\vspace{-0.5cm}
\caption{Study of the $\phi\to\pi^{0}\pi^{0}\gamma$ events:
a. $\chi^{2}$ distribution for a five photon sample after the 
constrained fit.
Histograms present the normalized simulated contribution from 
$\eta\gamma$ events (lower) and a sum of this distribution with 
that for $\omega\pi^{0}$ events (upper); 
b. $\pi^{0}\pi^{0}$ invariant mass for events with $2.4< \chi^{2}<6$. 
The histogram is simulation of $\eta\gamma$ events ;
c. $\pi^{0}\gamma$ invariant mass. Histograms are as in ``a'';
d. $\pi^{0}\pi^{0}$ invariant mass for events with $\chi^{2} <2.4$.
Histograms show a contribution from $\omega\pi^{0}$ events after the 
``anti-$\omega$'' cut (dashed) and a sum of this contribution
with  $\eta\gamma$ background (dotted).
}
\label{p0p0g}
\end{figure}

Candidates for this decay were selected from a sample of purely neutral events 
with the following criteria:
\begin{enumerate}
\item{
There are five or more photons in the CsI and BGO calorimeters with
the total energy deposition $E_{tot} > 1.75\cdot E_{beam}$.
The Monte Carlo simulation of the $\pi^{0}\pi^{0}\gamma$ events 
well reproduces the resolution over  $E_{tot}$ and shows that 
97\% of signal events survive this cut.
A requirement of minimum three photons in the CsI calorimeter gives 
a high trigger efficiency~\cite{NT}.
}
\item{
All photons have polar angles in the range 0.6-2.54 radians and
have energy higher than 20 MeV.
}
\item{
To select $\pi^{0}\pi^{0}\gamma$ events a constrained fit 
requiring energy-momen\-tum 
conservation was performed finding 
two best combinations of photon pairs with $\pi^{0}$ masses.
The $\chi^{2}$ distribution after the constrained fit is shown
in Fig.~\ref{p0p0g}a.  
Events with $\chi^{2} < 6$ 
were taken for further analysis.
This cut suppresses to the 10$^{-3}$ level the background from
the $\phi \to K^0_{S}K^0_{L}$ decay mode when    
$K^0_{S}$ decays to $\pi^{0}\pi^{0}$ and  $K^0_{L}$ produces the 
fifth cluster in the calorimeter.
}
\item{
Photons from reconstructed $\pi^{0}$'s in the BGO calorimeter 
have energy higher than 40 MeV. This cut removed incorrectly
reconstructed events with photons from the beam background.
} 
\item{
A selection cut 
$
|(P_{\pi^{0}}^{1}-P_{\pi^{0}}^{2})/(P_{\pi^{0}}^{1}+P_{\pi^{0}}^{2})|<0.8,
$
where $P_{\pi^{0}}^{1}, P_{\pi^{0}}^{2}$ were pion momenta,
was applied to remove incorrect combinations in which a free photon of
low energy was used as a part of a reconstructed $\pi^{0}$. 
}
\end{enumerate}


%
\subsection*{Background Subtraction}
The main background to events of interest comes from the 
processes $\phi\to\eta\gamma$, 
$\eta\to\pi^{0}\pi^{0}\pi^{0}$ with two lost photons and from
$e^+ e^- \to\omega\pi^{0}$ with the 
$\omega\to\pi^{0}\gamma$ decay.

The subtraction of the background from $\eta\gamma$
events was performed according to the simulation.
The invariant mass of the 
$\pi^{0}\pi^{0}$ system for $2.4 < \chi^{2} < 6$ is shown in
Fig.~\ref{p0p0g}b 
and demonstrates good agreement of the observed background
spectrum with simulation. The normalization of observed background
events to the collected integrated luminosity gives
$N_{\phi}=(20.6\pm1.0)$  millions of $\phi$ decays in agreement with the 
$(18.8\pm0.9)$ millions obtained from the analysis of 
seven photon events from the $\phi\to\eta\gamma\to\pi^{0}\pi^{0}\pi^{0}\gamma$ 
decay~\cite{CMD9911}. The latter number was used for the normalization of 
signal events.      

 The invariant mass of the 
$\pi^{0}\gamma$ system for selected events is shown in Fig.~\ref{p0p0g}c and
demonstrates the presence of  events from $e^+ e^-\to\omega\pi^{0}$ process.  
The number of $\omega\pi^{0}$ events was found to be $506\pm28$ with the
$\pi^{0}\gamma$ mass of $(782.0\pm0.9)$~MeV close to the world average
value of the $\omega$ meson mass~\cite{pdg}.
These events were used to check the above cut efficiencies and
event selection criteria. By looking for the $\omega\pi^{0}$ signal in 
a sample of six and seven photon events it was confirmed
that 92\% of signal events had exactly five photons. This was
used for the efficiency correction.  
 The simulated detection efficiency 
for $e^+ e^- \to\omega\pi^{0}$ events was found to be $(14.4\pm0.1)\%$ . 
\begin{figure}[tbh]
\vspace{-0.5cm}
\begin{center}
\psfig{figure= 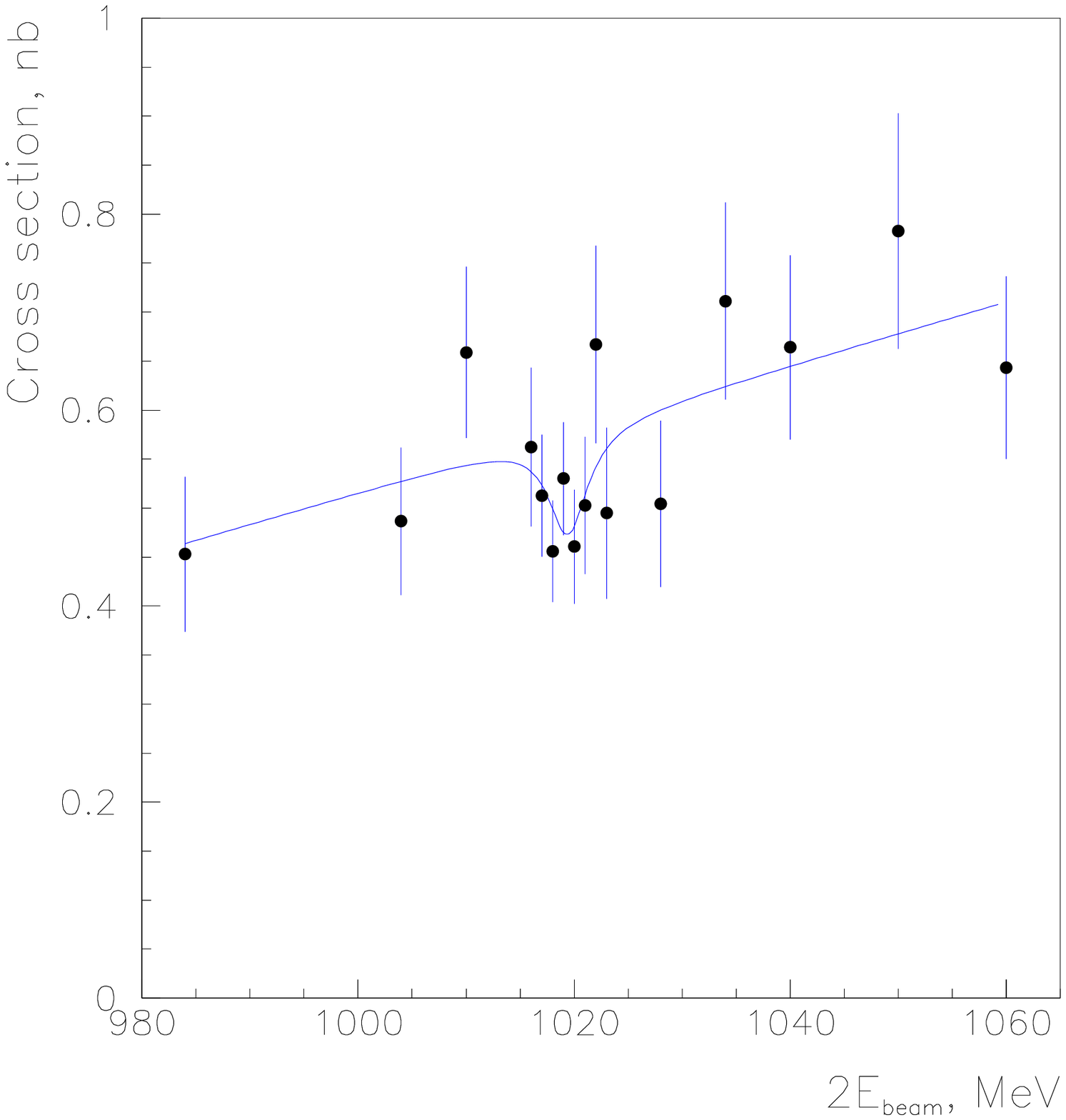, height=0.48\textwidth} 
\hfill
\psfig{figure= 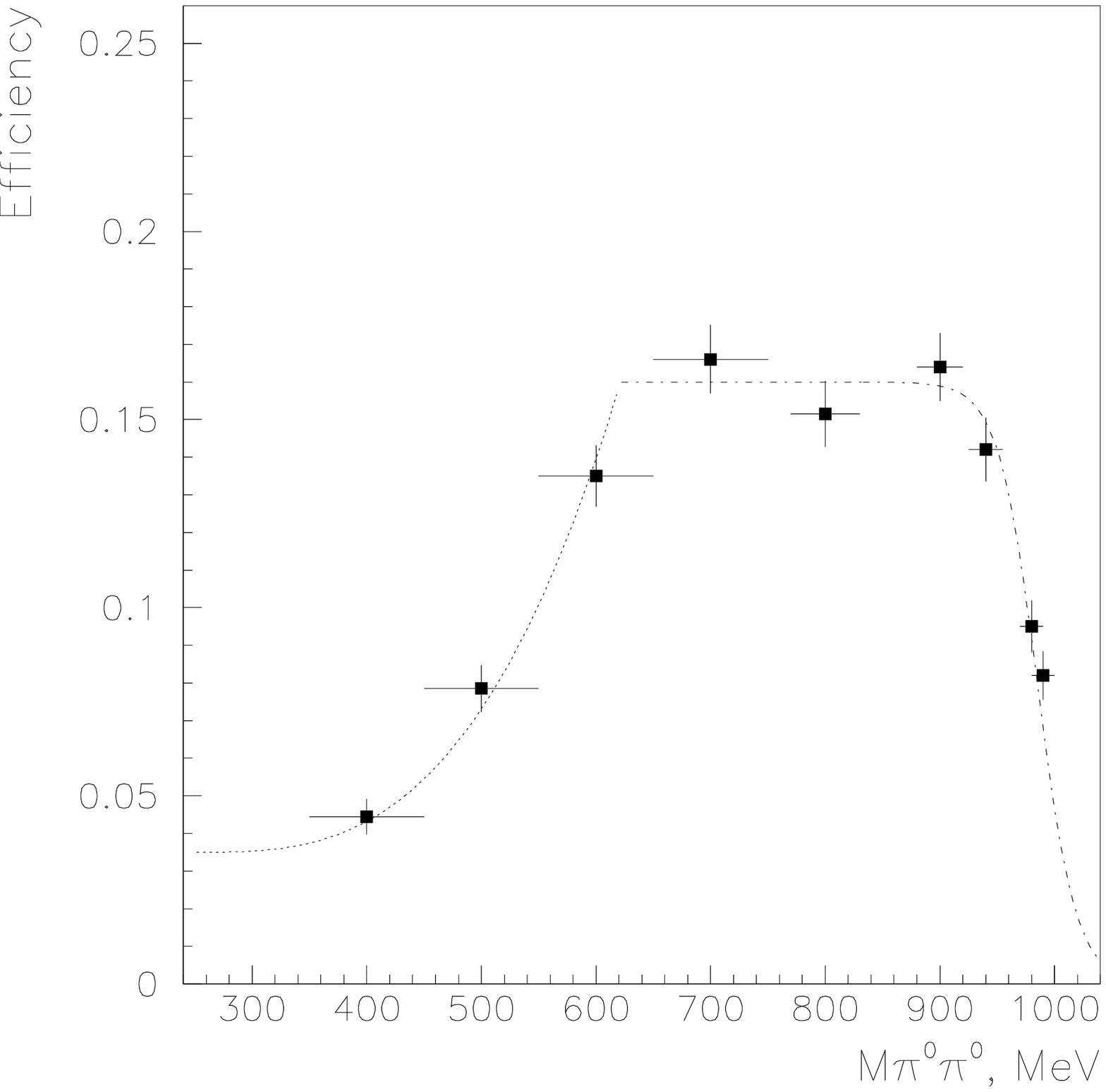, height=0.48\textwidth}
\parbox[t]{.45\textwidth}{
\caption{The cross section vs. energy for selected $\omega\pi^{0}$ events.}
\label{omega_pi0}}
\hfill
\parbox[t]{.45\textwidth}{
\caption{
The simulated efficiency vs. $\pi^{0}\pi^{0}$ invariant mass. 
The line is an approximation used for calculations.
}
\label{p0p0_eff}}
\end{center}
\end{figure}

The energy dependence of the cross section of the process
$e^+e^-\to\omega\pi^{0}\to\pi^{0}\pi^{0}\gamma$
is presented in Fig.~\ref{omega_pi0}. It was fit taking into account 
the interference of the non-resonant process with the $\phi$ decay.
The values of the        parameters of the 
 $\phi\to\omega\pi^{0}$ decay were taken from ~\cite{SND9865}.
 The non-resonant cross section obtained at 
$2E_{beam}=m_{\phi}$ is
 $\sigma_{0}=(0.58\pm 0.02\pm 0.04)$ nb consistent with 
that measured by SND~\cite{SND9865}. The second error represents
the systematic uncertainty caused by background subtraction.
 
The "anti-$\omega$" cut
M$_{inv}(\pi^{0}\gamma) < 750$ MeV reduces the admixture of
 $\omega\pi^{0}$ events to the level of about 5\% (see the dashed histogram in 
Fig.~\ref{p0p0g}d). 
However, it does not completely remove
 $\omega\pi^{0}$ events mostly because of
some incorrectly reconstructed events in which a free photon 
is used as a part of a
reconstructed $\pi^{0}$. These incorrect combinations 
were studied using simulated $\omega\pi^{0}$ events and experimental 
events from the ``off-$\phi$'' region where the process 
$e^+ e^- \to\omega\pi^{0}$ dominates. 
The ratio of the number of incorrectly reconstructed events to the total
number of  $\omega\pi^{0}$ events was found to be 0.060$\pm$0.003 for
simulation and 0.09$\pm$0.02 for experimental data. The difference was
used to estimate a systematic error
so that the final admixture of $\omega\pi^{0}$ events in the 
$\pi^{0}\pi^{0}\gamma$ sample was (5.0$\pm$1.5)\%.
\subsection*{Branching Ratio Calculation}
The invariant mass distribution of the
$\pi^{0}\pi^{0}$ system for $\chi^{2} < 2.4$ is shown in
Fig.~\ref{p0p0g}d 
with the expected backgrounds from $\omega\pi^{0}$ after the 
"anti-$\omega$" cut and from $\eta\gamma$ events. 
The distribution demonstrates the increase in the number of events with a 
high invariant mass (a free photon of low energy). In total,  
$268 \pm 27$ of $\pi^{0}\pi^{0}\gamma$ events have been found 
after all cuts and background subtraction. 

The detection efficiency as a function of  the $\pi^{0}\pi^{0}$
invariant mass was obtained using simulation of the process 
$e^+ e^- \to X(M)\gamma$ where X(M) was a particle with a small width and
variable mass M decaying into $\pi^{0}\pi^{0}$. 

The  angular 
distribution of the free photon was taken as 
$dN/d\theta_{\gamma}\approx (1+cos^{2}(\theta_{\gamma}))$.
The detection efficiency obtained 
is presented in Fig.~\ref{p0p0_eff} as a function of 
the $\pi^{0}\pi^{0}$ invariant mass.

\begin{figure}[tbhp]
\vspace{-1.0cm}
\psfig{figure= 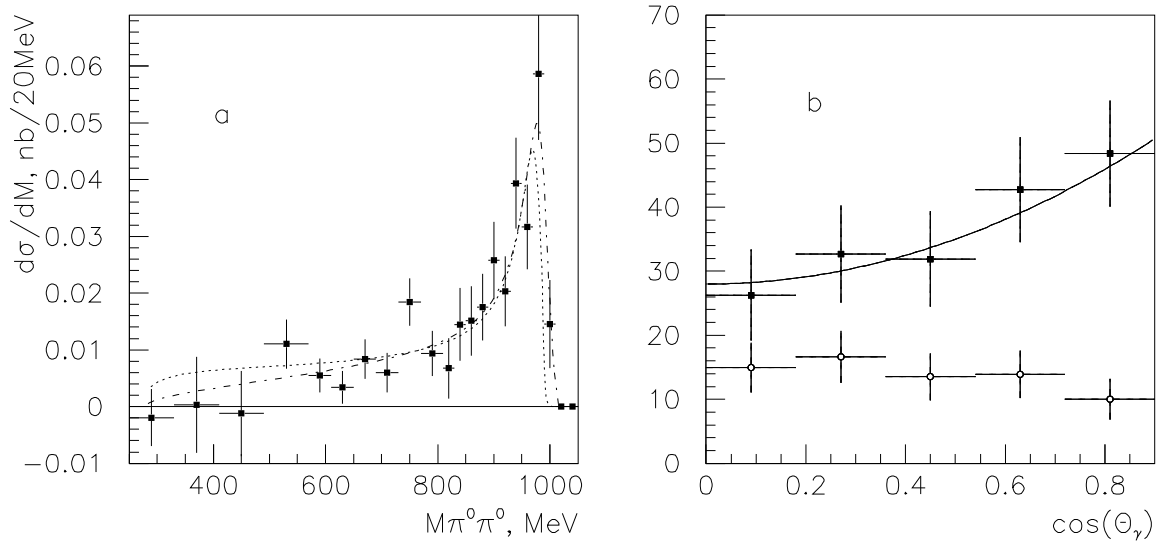, width=1.1\textwidth}
\caption{
a. Differential cross section vs invariant mass.
The dotted line is the  
four quark model fit with the  branching ratio of
$3.11\times 10^{-4}$.
The dashed line is the narrow pole fit;
b. Angular distribution for signal events and background (open points). 
The line is $dN/d\theta_{\gamma}\approx (1 + cos^{2}(\theta_{\gamma}))$
}
\label{p0p0g_spec}
\end{figure}

The $\pi^{0}\pi^{0}$ mass spectrum was obtained from the experimental 
distribution (Fig.~\ref{p0p0g}d)  after background subtraction and taking 
into account the detection efficiency for each histogram bin. The 
spectrum was normalized to the number of $\phi$ decays obtained from
the $\phi\to\eta\gamma\to\pi^{0}\pi^{0}\pi^{0}\gamma$ analysis~\cite{CMD9911}. 
The resulting differential cross section vs. invariant
mass is presented in Fig.~\ref{p0p0g_spec}a 
showing a resonance increase at high masses. 
Figure~\ref{p0p0g_spec}b presents the angular distribution for free 
photons for signal events with M$_{\pi^{0}\pi^{0}}>$ 800 MeV. The line 
shows the distribution expected for a scalar intermediate resonance:
 $dN/d\theta_{\gamma} \approx (1+cos^{2}(\theta_{\gamma}))$.  Open points 
show the angular distribution of the  subtracted background.
The branching ratio calculated from the integral over the differential
spectrum was found to be:

$Br(\phi\rightarrow\pi^{0}\pi^{0} \gamma)=(1.08\pm0.17\pm0.09)\times10^{-4}$\\
for the whole invariant mass range. The systematic error comes from the 
uncertainty of the background
subtraction (about 5\%) and from the uncertainty of the number of $\phi$
mesons (about 5\%). The main contribution to the statistical error comes 
from the region M$_{\pi^{0}\pi^{0}}<$ 550 MeV because of the big uncertainty 
of the background subtraction. 
 
This result can also be presented as:

$Br(\phi\rightarrow\pi^{0}\pi^{0} \gamma)
=(1.06\pm0.09\pm0.06)\times10^{-4}$ for M$_{\pi^{0}\pi^{0}}>$ 550
MeV;

$Br(\phi\rightarrow\pi^{0}\pi^{0} \gamma)
=(0.92\pm0.08\pm0.06)\times10^{-4}$ for M$_{\pi^{0}\pi^{0}}>$ 700
MeV;

$Br(\phi\rightarrow\pi^{0}\pi^{0} \gamma)
=(0.57\pm0.06\pm0.04)\times10^{-4}$ for M$_{\pi^{0}\pi^{0}}>$ 900
MeV.\\
The results above are 
consistent with those presented by 
SND~\cite{SND9865,SNDppg}.

 It should be mentioned that only results for  M$_{\pi^{0}\pi^{0}}>$ 700 MeV 
can be considered as model independent. For lower masses the
interference with  the nonresonant
($\omega\pi^{0}\to\pi^{0}\pi^{0} \gamma$)
and resonant ($\phi\to\rho\pi^{0}\to\pi^{0}\pi^{0}\gamma$) 
backgrounds can change the branching ratio. 
To extract these contributions a much better measurement at lower masses
is needed.

 About half of the signal events have the  free photon
energy below 100 MeV pointing to the presence of  the $f_{0}(980)$ 
intermediate state in the $\pi^{0}\pi^{0}$ system.
\section*{Data Interpretation}
For data interpretation,  similarly to our analysis of the 
$\phi\to\to\pi^{+}\pi^{-}\gamma$ decay~\cite{ppg},   
the model suggested in ~\cite{AchIvan,achap0p0} was used.
The model considers the $\phi\to f_{0}(980) \gamma$ decay 
under different assumptions about the 
$f_{0}$ meson structure (a two-quark or
four-quark state or $K\bar K$ molecule depending on the values of the coupling
constants g$^{2}_{K\bar{K}}/4\pi$, g$^{2}_{\pi\pi}/4\pi$) and calculates  
the differential cross section 
over the invariant mass of two pions $M_{\pi\pi}$.
Also considered are effects of other possible intermediate states like 
$\sigma\gamma$ and $\rho\pi^{0}\to\pi^{0}\pi^{0}\gamma$ with the
total contribution of about 15\%.  

The experimental spectrum Fig.~\ref{p0p0g_spec}a was fit using the
differential cross section from~\cite{AchIvan,achap0p0} 
assuming that the $\phi\to\pi^{0}\pi^{0}\gamma$ decay is completely 
dominated by the $f_{0}(980)\gamma$ mechanism.
The following model parameters have been obtained:
 $m_{f_{0}}=(977\pm 3\pm 6)$ MeV,  g$^{2}_{K\bar{K}}/4\pi =(2.44\pm
0.73)$~GeV $^2$,  g$^{2}_{\pi\pi}/4\pi = (0.40\pm0.06)$~GeV$^2$,
$Br(\phi\to f_{0}(980) \gamma) = (3.05\pm0.25\pm0.72)\cdot10^{-4}$.\\
The second error includes experimental uncertainties at high $\pi^{0}\pi^{0}$
masses and the 15\% effect of other processes.
The parameters obtained are in good agreement with those presented by the
SND group~\cite{SNDppg} and can be compared to the values obtained from 
our analysis of the $\phi\to\pi^+\pi^-\gamma$ channel~\cite{ppg}: 
$m_{f_{0}}=(976\pm 5\pm 6)$ MeV,
$Br(\phi\to f_{0}(980) \gamma) = (1.93\pm0.46\pm0.50)\cdot10^{-4}$.

The results for the $m_{f_{0}}$ and $Br(\phi\to f_{0}(980) \gamma)$ 
obtained from the $\pi^{0}\pi^{0}\gamma$ and $\pi^{+}\pi^{-}\gamma$  
final states are consistent within the errors. Therefore, we can perform
their combined analysis. A simultaneous fit of the photon spectra  
using the formulae from ~\cite{achap0p0,acha97,acha98} 
allowed to determine the model parameters with a better statistical
accuracy:\\
 $m_{f_{0}}=(975\pm4\pm6)$ MeV,  g$^{2}_{K\bar{K}}/4\pi =(1.48\pm
0.32)$~GeV $^2$,  g$^{2}_{\pi\pi}/4\pi = (0.41\pm0.06)$~GeV$^2$,\\
$Br(\phi\to f_{0}(980) \gamma) = (2.90\pm0.21\pm0.65)\cdot10^{-4}$, and
$\Psi = (1.47 \pm 0.19)$ radians. The $\chi^2$/d.f.=1.5 has been
obtained. The results of the fit are shown in Fig.~\ref{p0p0g_spec}a
by the dotted line. 
According to the model, the obtained values of the coupling constants 
(or the branching ratio) could only be
explained if $f_{0}$ is a four quark state.
The ratio R=g$^2_{K\bar{K}}$/g$^2_{\pi\pi}$ had weak
dependence on the $f_{0}$ structure and was found to be R=3.61$\pm$0.62.

In the above interpretation the whole visible signal was due to 
the $\phi\to f_{0}\gamma$ decay whereas
the influence of other possible mechanisms ($\rho\pi,\sigma\gamma$) 
was estimated to be about 15\% and included 
into the systematic error. 

 To understand the sensitivity of the  
branching ratio of the $\phi\to f_{0}(980)\gamma$ decay to the 
particular model, we
used another approach and performed  the
narrow pole fit. The $\pi^{0}\pi^{0}$ mass spectrum was fit
with the function:\\

$
\frac{d\sigma}{dM_{\pi\pi}}\sim 2M_{\pi\pi}\cdot
(1-\frac{M_{\pi\pi}^{2}}{4E^{2}_{beam}})\cdot
\sqrt{1-\frac{4m^{2}_{\pi}}{M_{\pi\pi}^{2}}}\cdot
|\frac{m_{f(980)}\Gamma_{f(980)}}{\Delta_{f(980)}}
+\frac{Ae^{i\psi}m_{f(1200)}\Gamma_{f(1200)}}{\Delta_{f(1200)}}|^{2}
$, \\

\noindent
where $\Delta_{f} = M_{\pi\pi}^{2}-m^{2}_{f}+iM_{\pi\pi}\Gamma(M_{\pi\pi})$.
The parameters of the $f_{0}(1200)$ (or $\sigma$) resonance could not be
extracted from our data and were fixed at 
$m_{f(1200)}=1200$ MeV and $\Gamma_{f(1200)} =600$ MeV.
These parameters have very small influence on the result and instead of
the $f_{0}(1200)$  Breit-Wigner the constant background amplitude 
$Ae^{i\psi}$ can be used. 
The relative phase $\psi$ was found to be close to zero and fixed at
that value. 

The following parameters have been obtained:\\
 $m_{f_{0}}=(987\pm7)$ MeV/c$^2$, $\Gamma_{f_{0}}=(56\pm20)$ MeV,
the relative amplitude A=0.22$\pm$0.09 (or
A=0.08$\pm$0.03 in case of constant background), and $\chi^2$/d.f.=0.96. 
The result of the fit is shown in
 Fig.~\ref{p0p0g_spec}a by the dashed-dotted line.

The corresponding branching ratio (assuming that only 1/3 is seen in the 
$\pi^{0}\pi^{0}$ mode) 
$Br(\phi \to f_{0}(980) \gamma) = (1.5\pm0.5)\cdot10^{-4}$ has been
found. 

The discussion above shows that the data obtained cannot be interpreted without
a resonance in the two pion mass spectrum at about 980 MeV
both in charged and neutral modes and the 
$\phi\to f_{0}\gamma$ branching ratio cannot be lower than about 
$1.0\cdot10^{-4}$.   
As many authors agreed~\cite{somef0}, a value that high can hardly 
be explained in
the frame of the two quark model of the $f_{0}(980)$ which 
predicts the branching ratio at the level of $0.5\cdot10^{-4}$. 
To study the influence of
other intermediate mechanisms, 
better measurements of the $\pi\pi$ mass spectra are needed. 
\section*{Search for {\boldmath $\phi\to\eta\pi^{0}\gamma$} Decay}
Using the five photon final state, one can also look for the decay mode
$\phi\to\eta\pi^{0}\gamma$ appearing when
$\eta$ decays into two photons. This analysis was based on  
the same event sample as in the study of the  
$\phi\to\pi^{0}\pi^{0}\gamma$ decay mode  
with the selection criteria
described above.

 A fit 
finding the best combination of photon pairs with the $\pi^{0}$ mass 
constrained and reconstructed
pion momentum less than 350 MeV/c was used. 
An additional requirement is that the difference in 
the energy of the photons from the found $\pi^{0}$ is less than
80\%. It rejects background from low
energy photons and increases the efficiency for events with a low energy free
photon because of a smaller probability of wrong combinations. 

\begin{figure}[tbh]
\vspace{-0.5cm}
\psfig{figure= 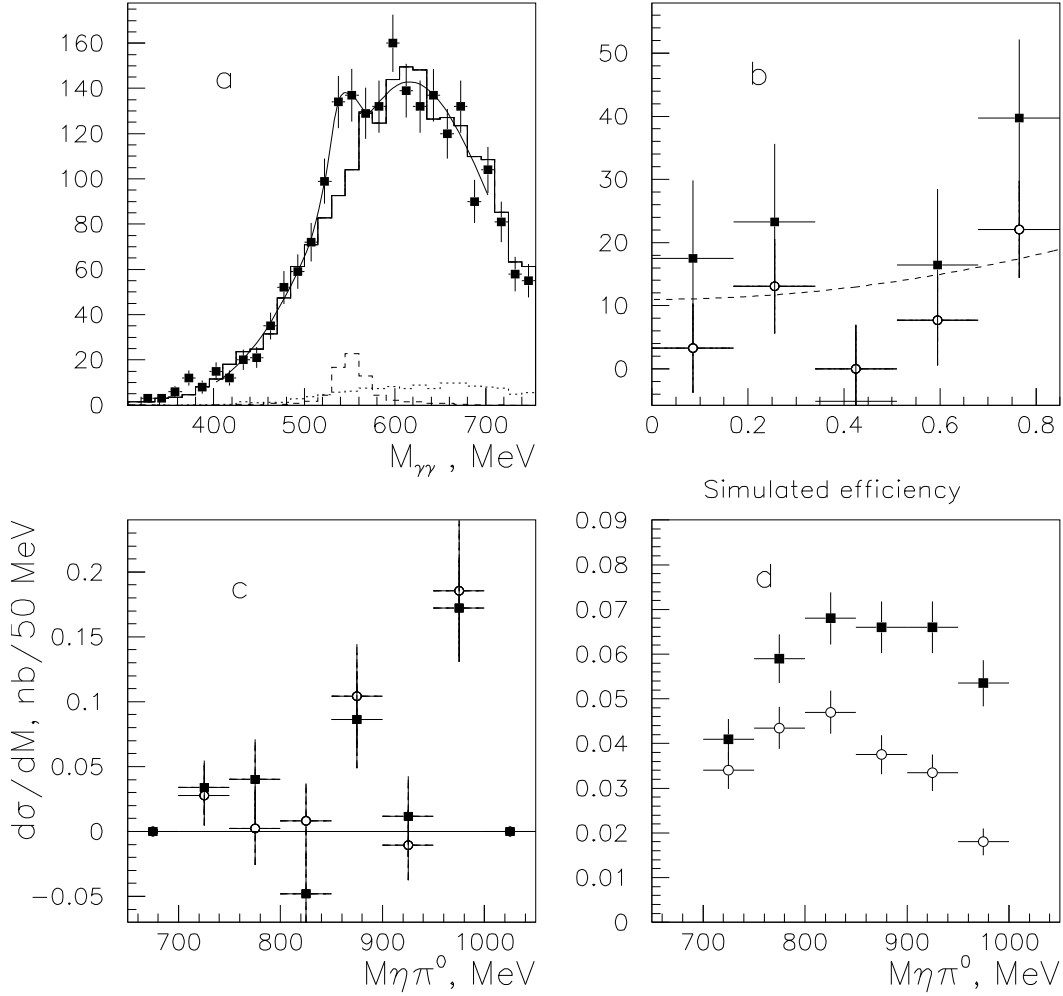, width=1.0\textwidth, height=0.9\textwidth}
\vspace{-0.8cm}
\caption{The $\phi\to\eta\pi^{0}\gamma$ study.
Open points correspond to the "strong" cuts:
a. Invariant masses of two photons with the highest energy for data (points) 
and simulation (histograms);
b. $cos \theta_{\gamma}$ distribution for events around the 
$\eta$ mass after background
subtraction. The line corresponds to the expected distribution
$dN/d\theta_{\gamma}\approx (1+cos^{2}(\theta_{\gamma}))$;
c. $\eta\pi^{0}$ mass distribution;
d. Simulated detection efficiency vs. invariant mass.
}
\label{etap0g}
\end{figure}

The invariant masses of 
the remaining most energetic photon pairs
are shown in Fig.~\ref{etap0g}a.

The main source of background to observed events is the decay
$\phi\to\eta\gamma\to 3\pi^{0}\gamma$ with two photons lost.
The cut $760< M_{\pi^{0}\gamma} < 805$ MeV from the 
$\pi^{0}\pi^{0}\gamma$
reconstruction almost removes the background from $\omega\pi^{0}$. 
The $\omega\pi^{0}$ events
remaining after the ``anti-$\omega$'' cut 
are shown by the dotted histogram.   

Over the broad background distribution 
the excess of $80 \pm 22$ events is observed at 
$(545 \pm 4)$ MeV compatible with the mass of the $\eta$ meson.
The  solid histogram shows the sum of the simulated 
background from $\phi\to\eta\gamma\to 3\pi^{0}\gamma$ 
and remaining $\omega\pi^{0}$ events normalized  
to the number of $\phi$ decays. These simulated events were used for
background subtraction.
The dashed histogram at 550 MeV shows  a simulated signal 
from the $\phi\to\eta\pi^{0}\gamma$ decay at the $1\times 10^{-4}$ level. 

 The distributions over $cos \theta_{\gamma}$ and invariant mass 
$M_{\eta\pi^{0}}$ are shown in Figs.~\ref{etap0g}b,c for events with
$510< M_{\gamma\gamma} < 590$ MeV after background subtraction and
taking into account 
the detection efficiency obtained by simulation (Fig.~\ref{etap0g}d). 
The following branching ratio:\\
 $Br(\phi\rightarrow\eta\pi^{0}\gamma)=(0.90\pm0.24\pm0.10)\times10^{-4}$\\
has been obtained. 
The systematic error comes from  the uncertainty of the
background subtraction and of the number of $\phi$'s taken for
normalization.

To check the stability of the result
 these distributions were also obtained in the case
of "strong" cuts when the additional requirement for
the reconstructed photon with the highest energy was applied: 
$E_{\gamma max}/E_{beam}< 0.75$. This cut reduced
the main background from $\phi\to\eta\gamma\to 3\pi^{0}\gamma$ by a factor
of 4, but at the same time considerably reduced the detection efficiency 
at higher
 $\eta\pi^{0}$ invariant masses (open points in Fig.~\ref{etap0g}d).  
The number of observed signal events dropped to $37 \pm 12$ and the
 obtained branching ratio was 
the same within statistical errors. 

The obtained invariant mass distribution shows the growth 
of the cross section to higher masses supporting the hypothesis
about the $a_{0}$(980) intermediate state. The obtained value of the 
branching ratio can be explained in the four quark model suggested in
~\cite{AchIvan,achap0p0}. 
\section*{ \boldmath Search for  $\eta\to\pi\pi$ Decays}
\hspace*{\parindent}
The selected  $\pi^{0}\pi^{0}\gamma$
events can be used to search for the
P- and CP-violating decay  $\eta\rightarrow\pi^0\pi^0$, where the 
 $\eta$ comes from
the radiative  $\phi\to\eta\gamma$ decay. 
From 18.8 million $\phi$ decays used 
for  the analysis of the $\pi^{0}\pi^{0}\gamma$ channel
one could expect about 236,000 events 
decayed via the  $\eta\gamma$ channel.
Such a P- and CP-violating decay should  be observed as a   
peak  in the invariant mass of two pions at  $M_{\pi\pi}=m_{\eta}$.
%
\begin {figure}[tbh]
\epsfig{figure= 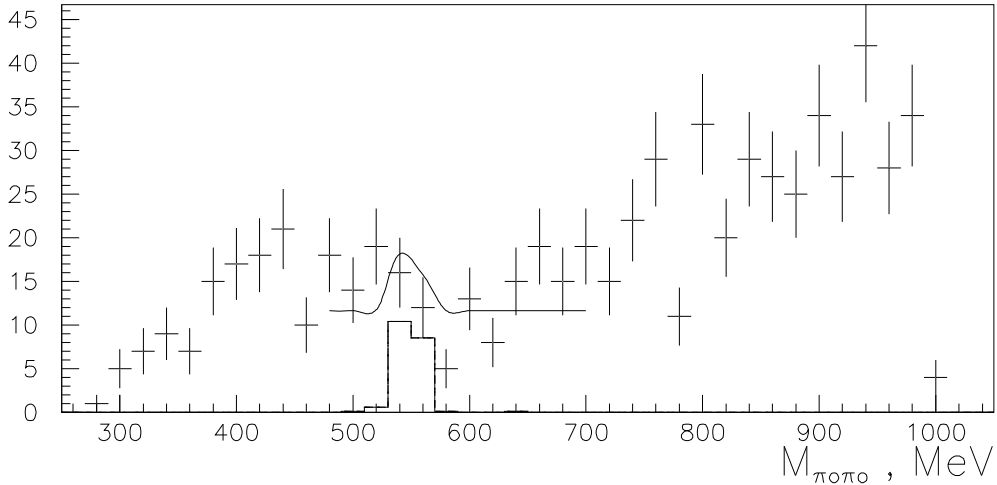, width=1.0\textwidth}
\caption{
Search for  $\eta\to\pi^{0}\pi^{0}$ decay. 
The histogram is simulation. The line shows a possible signal at 90\% CL.
} 
\label{etapipi}
\end{figure} 
Figure~\ref{etapipi} shows the experimental $\pi^{0}\pi^{0}$  mass
distribution from selected $\pi^{0}\pi^{0}\gamma$ events.
The line corresponds to a fit with a linear function and gaussian
distribution representing a possible signal at 90\% CL. It was found that 
a possible signal does not exceed  11 events for neutral decay mode. 
The histogram shows a simulated signal from  
the $\eta\to\pi^{0}\pi^{0}$ decay at 90\% CL. 
The detection efficiency found by simulation
was 0.108.
 The following result has been obtained:

Br($\eta\rightarrow\pi^{0} \pi^{0}) < 4.3\times 10^ {-4}$,\\
which should be compared to the best previous limit
$7\times10^{-4}$ for the neutral decay mode~\cite{SNDetap0p0}.

\section*{ \boldmath Conclusions}
\hspace*{\parindent}
Using 12.8 pb$^{-1}$ of data collected around
the   $\phi$ meson (about 19 million $\phi$ decays)  
events with five photons were selected.
The reconstruction of
the  $\pi^{0}\pi^{0}\gamma$ and $\eta\pi^{0}\gamma$ final states gives
the following model independent results:

 $Br(\phi\to\pi^{0}\pi^{0}\gamma)=(0.92\pm0.08\pm0.06)\times 10^{-4}$
 for M$_{\pi^{0}\pi^{0}}>$ 700 MeV,

 $Br(\phi\to\eta\pi^{0}\gamma)=(0.90\pm0.24\pm0.10)\times 10^{-4}$,\\
in agreement with those reported by the 
SND group~\cite{SNDppg,SNDa0g}.
A resonance at 980 MeV is observed in the $\pi^{0}\pi^{0}$ mass spectrum.
 
For the  $\pi^{0}\pi^{0}\gamma$ channel
 the branching ratio obtained by integration over the whole mass spectrum
was found to be

  $Br(\phi\to\pi^{0}\pi^{0}\gamma) = (1.08\pm 0.17\pm0.09)\times10^{-4}$,\\
 but this result can be affected by the possible interference with 
  $\omega\pi^{0}\to\pi^{0}\pi^{0}\gamma$ or
  $\rho\pi^{0}\to\pi^{0}\pi^{0}\gamma$ intermediate states at
 M$_{\pi^{0}\pi^{0}}<$ 700 MeV.

In the  $\pi\pi\gamma$ channel the $\phi\to f_{0}(980)\gamma$
mechanism dominates and the most consistent description for both charged and
neutral modes can be obtained in the four quark model with the branching
ratio

 $Br(\phi\to f_{0}(980)\gamma)=(2.90\pm0.21\pm0.65)\times 10^{-4}$.\\
However, this result is model dependent and it is shown that 
in another approach the branching ratio can decrease to about  
$1.5\times 10^{-4}$, still higher than the prediction of the two 
quark model. The difference between two approaches 
can be used as an estimate of the additional systematic uncertainty 
because of the model so that

 $Br(\phi\to f_{0}(980)\gamma)=(2.90\pm0.21\pm1.54)\times 10^{-4}$.

From the discussed models the following  $f_{0}(980)$ parameters
can be obtained:

  $m_{f_{0}} = (978\pm4\pm8)$ MeV, $\Gamma_{f_{0}} =(56\pm20\pm10)$ MeV,\\
where the second error represents a model dependent uncertainty.

In the   $\eta\pi^{0}\gamma$ final state events with the 
high  $\eta\pi^{0}$ invariant mass dominate. This supports the hypothesis 
about the  $a_{0}(980)$ intermediate state.  The obtained branching ratio 
is also higher than the prediction of the two quark model.  

For the P- and CP-violating 
decay of  $\eta\rightarrow\pi^{0}\pi^{0}$ 
the following upper limit at 90\% CL has been obtained:

  $Br(\eta\rightarrow\pi^{0}\pi^{0})~<~4.3\times10^{-4}$.\\
This result is the most stringent upper limit today.

The values of the branching ratios  $Br(\phi \to f_0(980)\gamma)$ and
$Br(\phi \to a_0(980)\gamma)$ obtained in this work are rather high and
according to theoretical predictions favour the four-quark structure
of these states~\cite{AchIvan,somef0}. However, predictions 
for the branching ratios significantly differ from one work to another 
and can hardly be considered as very accurate. Moreover, recently 
there were successful attempts to explain the observed distributions
in radiative decays  $\phi \to \pi \pi \gamma$ as well as their 
branching ratios without assumptions about the exotic structure of
the $f_0(980)$ meson~\cite{delb,marco}.

In general, the situation with the identification of the members of the
lowest scalar nonet and the related problem of the $f_0(980)$ and
$a_0(980)$ structure remains to be unclear and controversial. While
Ref.~\cite{ufn} presents many arguments in favour of the four-quark
structure of these states, other possibilities are widely discussed.
Ref.~\cite{locher} advocates the idea that the $f_0(980)$ meson doesn't
belong to the $q\bar{q}$ family and originates from the $K\bar{K}$ 
molecule embedded in the $\pi\pi$ continuum. However, in \cite{kim}
the idea of the loosely bound $K\bar{K}$ molecule is strongly disfavoured
and the possibility is discussed that both $f_0(980)$ and
$a_0(980)$ are conventional  $q\bar{q}$ states with properties strongly
distorted by coupling to the nearby  $K\bar{K}$ threshold. The small
value of the two-photon width of these states which traditionally
used to be one of the 
strong arguments in favour of their exotic structure
seems to be well accounted for in different theoretical models 
assuming the conventional $q\bar{q}$ structure of  $f_0(980)$ and $a_0(980)$
mesons \cite{delb,lucio,anis}. Note also the results of
OPAL \cite{opal} and DELPHI \cite{delphi} who studied the inclusive
production of  $f_0(980)$ and $a_0(980)$ mesons in hadronic $Z^0$
decays and found that their properties are consistent with the
normal  $q\bar{q}$ mesons. 

One can conclude that the problem of the  $f_0(980)$ and $a_0(980)$
structure is far from being solved and requires both new theoretical 
approaches and reliable, model independent experimental results.

\subsection*{Acknowledgements}
\hspace*{\parindent}
The authors are grateful to N.N.Achasov,  V.P.Druzhinin, 
V.V.Gubin, V.N.Ivanchenko  and A.I.Mil\-stein for useful
discussions and help with the data interpretation.
\end{document}